\documentclass{article}
\pdfoutput=1

\usepackage{algorithm}
\usepackage{algorithmicx}
\usepackage[noend]{algpseudocode}
\usepackage{amssymb}
\usepackage{amsmath}
\usepackage{caption}
\usepackage{subcaption}
\usepackage{listings}
\usepackage[numbers,sort&compress]{natbib}
\usepackage{siunitx}
\usepackage{booktabs}
\usepackage{tikz}
\usepackage{tikz-cd}

\DeclareMathOperator*{\argmax}{argmax}

\DeclareMathOperator*{\orient}{orient}

\usetikzlibrary{graphs}
\usetikzlibrary{quotes}
\usetikzlibrary{decorations.pathreplacing}
\usetikzlibrary{patterns}

\newcommand{\vqhull}{VQhull}
\newcommand{\qhull}{Quickhull}

\author{Thomas Koopman \and Jordy Aaldering \and Bernard van Gastel \and
Sven-Bodo Scholz}

\title{VQhull: a Fast Planar Quickhull}

\begin{document}

\maketitle

\begin{abstract}
Finding the convex hull is a fundamental problem in computational geometry.
Quickhull is a fast algorithm for finding convex hulls. In this paper, we
present VQhull, a fast parallel implementation of Quickhull that exploits vector
instructions, and coordinates CPU cores in a way that minimizes data movement.
This implementation obtains a sequential runtime improvement of
\qtyrange{1.6}{16}{\times}, and a parallel runtime improvement of 
\qtyrange{1.5}{11}{\times} compared to
the state of the art on the Problem Based Benchmark Suite. VQhull achieves
\qtyrange{85}{100}{\percent} of non-NUMA architectures' peak bandwidth, 
and \qtyrange{66}{78}{\percent} on our two-CPU NUMA system.
This leaves little room for further improvements.

A $4\times$ speedup on $8$ cores has a parallel efficiency of 
\qty{50}{\percent}. This suggests a waste of energy, but our
measurements show a more complicated picture: energy usage may even be lower in 
parallel. Quickhull serves as a case study that runtime and energy consumption 
do not go hand in hand.
\end{abstract}

\section{Introduction}

The convex hull of a set of points is the smallest superset closed under taking
convex combinations, a type of weighted average. The convex hull of a finite set
in 2D is a polygon, which can be described by its vertices. Finding these
vertices and listing them in clockwise order is a fundamental algorithm in
computational geometry. Some problems make use of the convex hull construction
directly, while others use a convex hull algorithm as subroutine. Applications
can be found in a wide range of domains, such as collision
detection~\cite{Tang08}, path planning~\cite{Meeran97}, and urban
planning~\cite{Kang17}.

There exists a multitude of algorithms for finding the convex
hull~\cite{Graham72, Jarvis73, Eddy77, Preparata77, Bykat78, Akl78, Andrew79,
Clarkson93, Barber96, Chan96}. An implementation of \qhull{} is faster than
other algorithms on most data sets, but relating the runtime performance to the
peak performance of the hardware shows there is still significant room for
improvement. In this paper we aim to close this gap by providing a novel
parallelisation of the \qhull{} algorithm called \vqhull{}.

Obtaining good performance on a multicore machine is a matter of exposing
sufficient parallelism and minimising data movement. \vqhull{} uses parallelism
by both vector instructions and multiple CPU cores. We need to make minor
algorithmic adjustments to make effective use of the memory subsystem.

The ICT sector is currently estimated to account for
\qtyrange{2.1}{3.9}{\percent} of global emissions ---~more than the aviation
industry~\cite{freitag2021real}. While performance analyses usually focus only
on runtime, it is clear that reducing the energy consumption of software is
crucial for reducing greenhouse gas emissions of ICT. We describe our method for
measuring energy consumption, using which we evaluate the energy-efficiency of
our approach.

We make the following contributions.
\begin{itemize}
    \item We present a vectorized algorithm for extracting two disjoint subsets
    in-place. (Section~\ref{sec:subset})
    \item We implement \vqhull{} with a parallelized subset extraction over
    multiple cores in a bandwidth-friendly manner. (Section~\ref{sec:impl})
    \item We provide an extensive runtime performance and energy consumption
    analysis of our approach compared to the state of the art, on three
    platforms. (Section~\ref{sec:evaluation})
\end{itemize}

\section{Background}

\subsection{\qhull{} Algorithm}\label{sec:preliminaries}

The \qhull{} algorithm~\cite{Barber96} is a widely used technique for finding
the convex hull of a finite point set. It operates similarly to the QuickSort
algorithm, as both apply a divide and conquer approach to recursively partition
a set.

\begin{figure}[ht]
    \centering
    \includegraphics[width=0.5\linewidth]{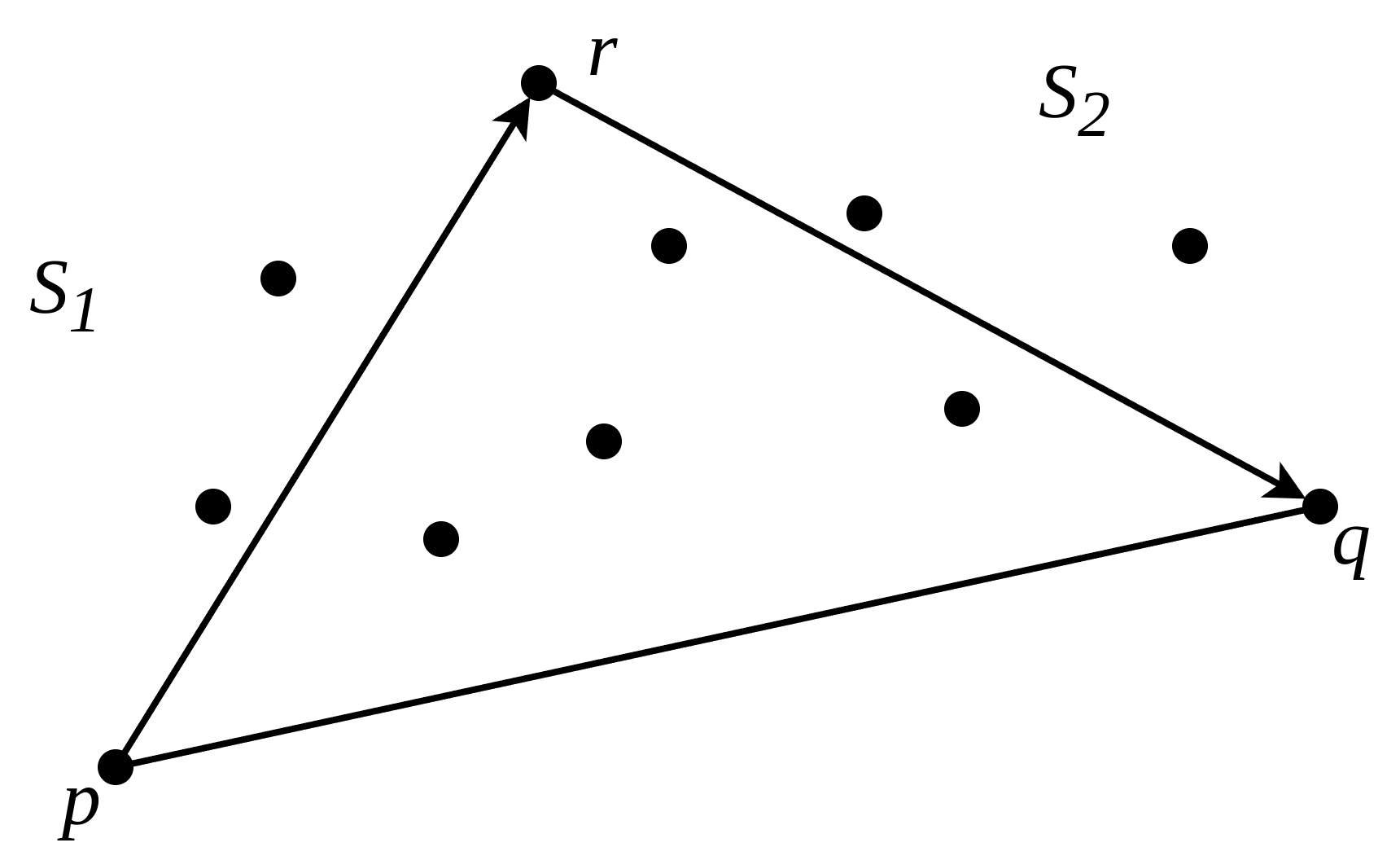}
    \caption{An example of the first partitioning step of the \qhull{} algorithm.}
    \label{fig:quickhull_vqhull}
\end{figure}

Figure~\ref{fig:quickhull_vqhull} illustrates the basic idea behind
\qhull{}---Algorithm~\ref{alg:quickhull_vqhull}. If we have three points $p$, $r$, $q$
on the convex hull of $P$, then the points inside that triangle are not in the
convex hull and can be discarded. The points $S_1$ to the left of
$\overrightarrow{pr}$, and the points $S_2$ to the left of $\overrightarrow{rq}$
may be on the convex hull, and are recursively inspected.

To find three points on the convex hull, we can start with the leftmost and
rightmost points $p$ and $q$. We use the degenerate triangle $\Delta pqp$ to
bootstrap the algorithm. The arguments of the recursive function \texttt{Hull}
give two points on the convex hulls of $S_1$, $S_2$. To find the third point we
use the following observation. If $x$, $y$ is on $CH(P)$, then a point $z \in P$
with maximum distance to $\overrightarrow{xy}$ must also be on the convex hull.

\begin{algorithm}[ht]
    \begin{algorithmic}[1]
        \Require The set of points $P \subset \mathbb{R}^2$.
        \Ensure The vertices $H \subseteq P$ of the convex hull in clockwise order.
        \State Let $p \in P$ be the leftmost point.
        \State Let $q \in P$ be the rightmost point.
        \State $S_1 := \{u \in P \mid u \text{ to the left of } \overrightarrow{pq}\}$
        \State $r_1 := \argmax\limits_{u \in S_1} d(u, \overrightarrow{pq})$
        \State $S_2 := \{u \in P \mid u \text{ to the right of } \overrightarrow{pq}\}$
        \State $r_2 := \argmax\limits_{u \in S_2} d(u, \overrightarrow{pq})$
        \State $H := \{p\}$ $\cup$ $\{q\}$ $\cup$
                     \Call{Hull}{$S_1$, $p$, $r_1$, $q$} $\cup$
                     \Call{Hull}{$S_2$, $q$, $r_2$, $p$}
        \Function{Hull}{$P$, $p$, $r$, $q$}
            \If{$|P| \leq 1$}
                \State \Return{$P$}
            \Else{}
                \State $S_1 := \{u \in P \mid u \text{ to the left of } \overrightarrow{pr}\}$
                \State $r_1 := \argmax\limits_{u \in S_1} d(u, \overrightarrow{pr})$
                \State $S_2 := \{u \in P \mid u \text{ to the left of } \overrightarrow{rq}\}$
                \State $r_2 := \argmax\limits_{u \in S_2} d(u, \overrightarrow{rq})$
                \State \Return{$\{r\}$ $\cup$
                               \Call{Hull}{$S_1$, $p$, $r_1$, $r$} $\cup$
                               \Call{Hull}{$S_2$, $r$, $r_2$, $q$}}
            \EndIf
        \EndFunction
    \end{algorithmic}
    \captionof{algorithm}{Quickhull algorithm.}
    \label{alg:quickhull_vqhull}
\end{algorithm}

We can efficiently and accurately tell whether $u$ is to the left of
$\overrightarrow{pq}$ by testing
\[
    (p_x - u_x) \cdot (q_y - u_y) > (p_y - u_y) \cdot (q_x - u_x),
\]
and check whether $u$ is farther than $u'$ from $\overrightarrow{pq}$ by testing
\[
    (q_y - p_y) (u_x - u'_x) < (q_x - p_x) (u_y - u'_y).
\]

The partitioning step is then recursively applied in
Algorithm~\ref{alg:quickhull_vqhull} on the remaining subsets $S_1$ and $S_2$. Each
call adds the element of its subset with maximum distance to
$\overrightarrow{xy}$ to the convex hull.

\subsection{Performance on Central Processing Units}

Implementation-related performance improvement comes mainly from exposing
parallelism and reducing data movement. We review what this looks like on the
hardware we target, a Central Processing Unit (\textit{CPU}), so we can design
the parallel algorithm with this in mind.

\subsubsection{Branch Prediction}

A CPU has multiple execution units per core. If there are no dependencies
between instructions, these units can execute instructions in parallel
(\textit{instruction level parallelism}), benefiting the performance. To keep a
steady flow of instructions, the CPU will make a guess when encountering a
conditional jump---generated by if statements and loops in high level languages.
This is called branch prediction and improves performance when guessed
correctly. If the guess is incorrect, the CPU must make an expensive recovery.

\subsubsection{Vectorization}

A \textit{vector} or \textit{SIMD} instruction is one that operates on multiple
primitive types (such as integers or floating point numbers) at the same time.
The number of elements is called the \textit{vector width}, and is typically
between $2$ and $8$ elements.

These instructions can be used to do multiple arithmetic operations such as
\texttt{a[0] = b[0] + c[0]; a[1] = b[1] + c[1];} in one instruction, but also
data-movement such as \texttt{a[0] = flag[0] ? b[0] : c[0]; a[1] = flag[1] ?
b[1] : c[1];}. This reduces the number of instructions necessary, and
instructions that permute data can also be used instead of branches.

\subsubsection{Data Movement and Multithreading}

This parallelism increases the number of computations we can do in a given
interval, but that is not the only limiting factor. After all, we also need to
load and store data fast enough. We call applications limited by the number of
computations we can do \textit{compute-bound}, and applications limited by data
movement \textit{memory-bound}.

Branch prediction and vectorization happens on each core. Similarly, each core
also has some fast local memory called a cache. These resources scale with using
more cores. The main memory however, shares a common data bus with all cores.
For this reason problems are more likely to become memory-bound when using more
cores.

\section{Subset Extraction}\label{sec:subset}

The main computation of Algorithm~\ref{alg:quickhull_vqhull} is to extract subsets
$S_1$, $S_2$ from $P$. This resembles the partition step in QuickSort, except
that we discard more points. This makes it more like a Dutch National Flag (DNF)
problem, where we want to extract three sets: $S_1$, $S_2$, and $P \backslash
(S_1 \cup S_2)$. We can implement this DNF-like problem more efficiently by
exploiting that we do not need to preserve the last set. We first describe a
vectorized algorithm for a single core, and then parallelise it over multiple
cores. Though the algorithm on one core is parallel in its use of SIMD
instructions and out-of-order execution, we follow convention and call it
sequential.

\subsection{Sequential Extraction}

The main idea behind the extraction on a single CPU core is to classify multiple
points simultaneously by using vector instructions. These registers can hold
multiple elements, and corresponding instructions can operate on all of them
simultaneously.

The main instruction used is a compression, \texttt{vcompresspd}, from the Intel
AVX-512 instruction set, illustrated in Figure~\ref{fig:compress}. This
instruction takes a vector register of \texttt{double}s and an integer whose
bits encode which elements in the vector should be compressed. It then stores
these elements contiguously in memory. We can emulate \texttt{vcompresspd} on
CPUs with different instruction sets~\cite{Blacher22}. The compression can be
used to extract the subsets from a single vector register. We can use this
instruction twice to extract the elements of $S_1$ and $S_2$ from a vector
register.

\begin{figure}[ht]
    \resizebox{0.5 \linewidth}{!}{%
        \begin{tikzpicture}
            \foreach \i in {0, ..., 3} {
                \draw (\i, 0)  rectangle (\i + 1, 0.5)  node[midway] {$p_{\i}$};
            }
            \draw (0, 0.75)  rectangle (1, 1.25) node[midway] {$0$};
            \draw (1, 0.75)  rectangle (2, 1.25) node[midway] {$1$};
            \draw (2, 0.75)  rectangle (3, 1.25) node[midway] {$1$};
            \draw (3, 0.75)  rectangle (4, 1.25) node[midway] {$0$};

            \draw (0, -0.75)  rectangle (1, -0.25) node[midway] {$p_1$};
            \draw (1, -0.75)  rectangle (2, -0.25) node[midway] {$p_2$};

            \draw [->] (1.5, -0.05) -- (0.5, -0.2);
            \draw [->] (2.5, -0.05) -- (1.5, -0.2);

            \node at (-1, 1) {\texttt{mask}};
            \node at (-1, 0.25) {\texttt{vector}};
            \node at (-1, -0.5) {\texttt{memory}};
        \end{tikzpicture}
    }
    \caption{The \texttt{vcompresspd} instruction}
    \label{fig:compress}
\end{figure}

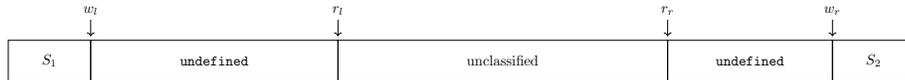
\begin{figure*}[ht]
    \resizebox{\linewidth}{!}{%
        \begin{tikzpicture}
            \draw (0, 0)  rectangle (2, 1)  node[midway] {$S_1$};
            \draw (2, 0)  rectangle (8, 1)  node[midway] {\texttt{undefined}};
            \draw (8, 0)  rectangle (16, 1) node[midway] {unclassified};
            \draw (16, 0) rectangle (20, 1) node[midway] {\texttt{undefined}};
            \draw (20, 0) rectangle (22, 1) node[midway] {$S_2$};

            \draw[->] (2,  1.5) node[above] {$w_l$} -- (2,  1.1);
            \draw[->] (8,  1.5) node[above] {$r_l$} -- (8,  1.1);
            \draw[->] (16, 1.5) node[above] {$r_r$} -- (16, 1.1);
            \draw[->] (20, 1.5) node[above] {$w_r$} -- (20, 1.1);
        \end{tikzpicture}
    }
    \caption{Invariant for extracting the subsets $S_1$ and $S_2$ from a set of points $P$.}
    \label{fig:invariant_qhull}
\end{figure*}

To extend the extraction to an entire array, we repeatedly use the compression
instruction while maintaining an invariant illustrated in
Figure~\ref{fig:invariant_qhull}. This is Algorithm~\ref{alg:subset}, which
maintains---for $d$ the number of elements per register---that
\begin{enumerate}
    \item we have $P[0, w_l) \subseteq S_1$,
    \item we have $P[w_r, |P|) \subseteq S_2$,
    \item either $r_l - w_l \geq d$ or $w_r - r_r \geq d$,
    \item the unclassified points are in $P[r_l, r_r)$ or buffered.
\end{enumerate}

\begin{algorithm}[ht]
    \begin{algorithmic}[1]
        \Require A set of $n$ points $P \subset \mathbb{R}^2$.
        \Ensure Pointers $w_l$, $w_r$ and a permutation of $P$ such that
        $P[0, w_l) = S_1$ and $P[w_r, |P|) = S_2$.
        \State $w_l = 0$; $r_l = d$; $r_r = |P| - d$; $w_r = |P|$;
        \State bufL = $P[0, d)$; bufR = $P[n - d, n)$;
        \While{$r_l < r_r$}
            \If{$w_l - r_l \geq r_r - w_r$}
                \State reg = $P[r_l, r_l + d)$
                \State $r_l = r_l + d$
            \Else
                \State $r_r = r_r - d$
                \State reg = $P[r_r, r_r + d)$
            \EndIf
            \State extract $S_1$ from reg, write to $w_l$, increase $w_l$
            \State extract $S_2$ from reg, decrease $w_r$, write to $w_r$
        \EndWhile
        \State extract $S_1$, $S_2$ from bufL and bufR
    \end{algorithmic}
    \captionof{algorithm}{Subset extraction.}
    \label{alg:subset}
\end{algorithm}

We establish the invariant with step 1 and 2 of Algorithm~\ref{alg:subset}.
These buffered elements can be safely overwritten. Our writes respect assertions
1 and 2. We read $d$ elements from the side of the array where the distance
between read and write pointer is smallest, ensuring that on that side no points
are overwritten. Assertion 3 ensures there are no points overwritten on the
other side either, which implies 4. Therefore, we only have to show that
assertion 3 holds. For this reason, we ensure at the initialization of the
algorithm that $(r_l - w_l) + (r_r - w_l) \geq 2d$. As we read $d$ elements at
each step, and write less than or equal to $d$ elements, this inequality is
maintained. The sum $(r_l - w_l) + (r_r - w_l)$ is at least $2d$, so the minimum
$\min(r_l - w_l, r_r - w_l)$ is at least $d$.

\subsection{Parallel Extraction}

Extracting $S_1$ and $S_2$ has linear complexity, so it is very likely to be
limited by bandwidth. For this reason, minimizing data movement is the primary
design goal of the parallel algorithm. The algorithm consists of two conceptual
steps: a parallel step where each thread independently works on their local part
of $P$, and a cleanup step that merges the subsets of each thread.

\paragraph{Parallel step}

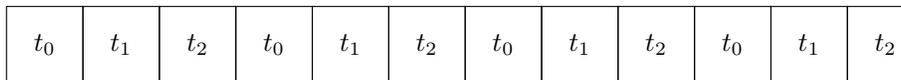
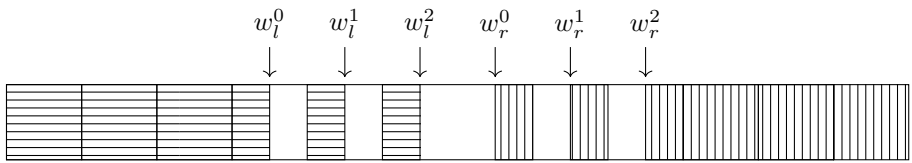
\begin{figure}[ht]
    \begin{subfigure}{\linewidth}
        \resizebox{\linewidth}{!}{%
            \begin{tikzpicture}
                \foreach \i in {0, ..., 10} {
                    \draw (\i, 0) rectangle (\i + 1, 1);
                }
                \draw (11, 0) rectangle (11.8, 1);

                \foreach \i in {0, ..., 3} {
                    \node at (3 * \i + 0.5, 0.5) {$t_0$};
                }
                \foreach \i in {0, ..., 3} {
                    \node at (3 * \i + 1.5, 0.5) {$t_1$};
                }
                \foreach \i in {0, ..., 3} {
                    \node at (3 * \i + 2.5, 0.5) {$t_2$};
                }
            \end{tikzpicture}
        }
        \subcaption{Block-cyclic distribution. Each square represents blocksize
        $b$ elements, except the last one that may contain less.}
        \label{fig:blockcycl:a}
    \end{subfigure}
    \begin{subfigure}{\linewidth}
        \resizebox{\linewidth}{!}{%
            \begin{tikzpicture}
                \draw (0, 0) rectangle (11, 1);

                \foreach \t in {0, ..., 2} {
                    \draw[pattern=horizontal lines, pattern color=black]
                        (3 * 0 + \t, 0) rectangle (3 * 0 + \t + 1, 1);
                    \draw[pattern=horizontal lines, pattern color=black]
                        (3 * 1 + \t, 0) rectangle (3 * 1 + \t + 0.5, 1);
                    \draw[->] (3 * 1 + \t + 0.5, 1.5) node[above] {$w_{l}^{\t}$}
                                -- (3 * 1 + \t + 0.5, 1.1);

                    \draw[->] (3 * 2 + \t + 0.5, 1.5) node[above] {$w_{r}^{\t}$}
                                -- (3 * 2 + \t + 0.5, 1.1);
                    \draw[pattern=vertical lines, pattern color=black]
                        (3 * 2 + \t + 0.5, 0) rectangle (3 * 2 + \t + 1, 1);
                    \draw[pattern=vertical lines, pattern color=black]
                        (3 * 3 + \t, 0) rectangle (3 * 3 + \t + 1, 1);
                }
            \end{tikzpicture}
        }
        \subcaption{Result of the parallel step. The horizontal lines indicate
        $S_1$, the vertical lines $S_2$, and the blank space is undefined. The
        superscript indicates which thread the write pointers belong to.}
        \label{fig:blockcycl:b}
    \end{subfigure}
    \caption{Parallel partition step for $3$ threads $t_0$, $t_1$, $t_2$}
    \label{fig:blockcycl}
\end{figure}

We first divide the points $P$ over all threads. As we would like to have most
points be in the correct position after the partitioning step, each thread
should have points on the left and right side of the array. This makes the
block-cyclic distribution, illustrated in Figure~\ref{fig:blockcycl:a} suitable.
To be more precise, the block-cyclic distribution over $T$ threads with block
parameter $b$ partitions an index set $[0, n)$ into $T$ index sets
\[
    I^{(t)} = \{(pk + s)b + j \mid 0 \leq j < b, \quad 0 \leq (pk + s)b + j < n\}
\]
Thread $t$ can then independently extract the subsets of $P^{(t)} := \{P_i \mid
i \in I^{(t)}\}$. Figure~\ref{fig:blockcycl:b} illustrates the resulting array
and write pointers.

The key observation is that for
\[ w_l^{\min} := \min_{t} w_l^t \]
\[ w_l^{\max} := \max_{t} w_l^t \]
\[ w_r^{\min} := \min_{t} w_r^t \]
\[ w_r^{\max} := \max_{t} w_r^t \]
all points in $P[0, w_l^{\min})$ are in $S_1$, all points in $P[0, w_r^{\max})$
are in $S_2$, and all points in $P[w_l^{\max}, w_r^{\min})$ are undefined. This
means only the points in $P[w_l^{\min}, w_l^{\max})$ and $P[w_r^{\min},
w_r^{\max})$ are potentially incorrect.

\paragraph{Cleanup}

For random input, or input that is already a convex hull, the write pointers of
different threads will be close to each other. This means only few points are
incorrect, so it is acceptable to classify the points in $P[w_l^{\min},
w_l^{\max}) \cup P[w_r^{\min}, w_r^{\max})$ sequentially. We use the Dutch
National Flag algorithm for this problem. A challenge for this algorithm is that
we need to test whether a point does not belong to $S_1$ or $S_2$. We overwrite
points, so the actual values of those points is undefined. The only way to test
this is by finding out to which thread the point belongs, and then checking
whether the index is between the write pointers of that thread. To make this
easier, each thread sets their undefined points to NaN.

\section{Implementation}\label{sec:impl}

We implement the algorithm using the Highway library~\cite{Blacher22} for
vectorization, and OpenMP for multithreading. Our library is available with
Fortran and C interfaces under a GPLv3 licence~\cite{implementation-vqhull}. The basic
idea is faithful to Algorithm~\ref{alg:quickhull_vqhull} and the subset extraction, but
several small changes are necessary to obtain good performance, which we
document here.

\paragraph{Parallelization}
We parallelize both the recursion and the subset extraction. We divide the
available threads over the recursive calls in the same ratio as $|S_1| : |S_2|$.
When we have only one thread left, we switch to the sequential implementation.

\paragraph{Bandwidth considerations}
To reduce the number of pas\-ses over the data, we compute the farthest points
$r_1$, $r_2$ in the same pass as the subset extraction.

Writes to and from main memory are done in aligned chunks, typically
\qty{64}{\byte}, called cachelines. This has two consequences we need to
mitigate. First, false sharing: two processors cannot write concurrently to the
same cacheline, as one processor would overwrite the result of the other. This
causes an expensive cacheline invalidation. Second, partial writes to a
cacheline still incur the full cost of writing to an entire cacheline.

To ensure each processor works on a disjoint cacheline, we pick the block size
$b$ equal to a multiple of the cacheline. We also align $P$ to cacheline
boundaries by processing a small number of elements at the start and end of the
array that do not fill an entire cacheline sequentially.

CPUs try to mitigate partial transactions to main memory by accumulating writes
to the same cacheline in a limited number of buffers. Writes to main memory are
delayed as long as possible, so more data can be sent to main memory in a single
transaction over the bus. On some of our test systems these buffers are
overwhelmed because we write in an irregular pattern to four different memory
regions ($x$- and $y$-coordinates from both the left and right side), and
because we do many writes of irregular length that may cross cacheline
boundaries. To give the CPU a better chance of combining bus traffic, we first
write to small circular buffers that fit in the first-level data cache. These
writes do not cause traffic to main memory. Once full, we can efficiently empty
the buffers to main memory by using vectorized stores to aligned addresses.

\section{Performance Evaluation}\label{sec:evaluation}

Comparing performance is only useful against the fastest available algorithm and
implementation. The original authors of Quickhull, the Computational Geometry
Algorithms Library \cite{CGAL}, and the Problem Based Benchmark Suite
\cite{pbbs} provide implementations of Quickhull, called Qhull, CGAL, PBBS
respectively. CGAL also provides implementations for some of the other
algorithms. Preliminary testing on a laptop (Table~\ref{tab:reference}), shows
that PBBS's implementation of \qhull{} outperforms its closest competitor by a
factor $2$, so that is what we use as baseline.

\begin{table}[ht]
    \centering
    \begin{tabular}{cc}\toprule
     Implementation     & Runtime \\\midrule
     PBBS               & 0.31    \\
     CGAL Quickhull     & 0.61    \\
     CGAL Akl-Toussaint & 0.60    \\
     CGAL Bykat         & 0.73    \\
     CGAL Eddy          & 0.98    \\
     Qhull              & 1.1     \\
     CGAL Graham        & 1.3     \\
     CGAL Jarvis        & 208     \\
    \bottomrule\end{tabular}
    \caption{Runtime in seconds for a disk of $10^7$ points, sequential
    implementation.}
    \label{tab:reference}
\end{table}

The Problem Based Benchmark Suite also describes three datasets which we use for
our evaluation. Each of these consists of $10^8$ points drawn from different
distributions.

\subsection{Evaluation Platforms}

We evaluate our implementation on three machines. The first, cn125, has powerful
vector instructions, a modest amount of cores and memory, and a simple memory
architecture. The second, cn132, has less powerful vector instructions and needs
to implement the compression instruction in software. It has two CPUs which
gives it more cores and memory bandwidth. One of these CPUs can access the
other's memory, but at a performance penalty. This is called \textit{non-uniform
memory access} (NUMA). For this reason, we ensure allocations are equally
distributed between the two CPUs by using \texttt{numactl --interleaved all}.
The third system, \textit{laptop} has the same instruction set as cn132, but is
a single consumer CPU. This CPU is a heterogeneous system with performance and
efficiency cores, but we only use the former by running the experiments under
\texttt{numactl -C 0-7}.

We summarize the systems in Table~\ref{tab:system}. There can be a significant
gap between the bandwidth the RAM is capable off, and what is achievable in
practice. The STREAM benchmark~\cite{McCalpin95} is a collection of simple
functions that can be used to measure what memory bandwidth is achievable in
practice. To most accurately mirror the access pattern of \vqhull{}, we report
the Scale benchmark function, which reads and writes to the same array.

\begin{table*}[ht]
    \centering
    \begin{tabular}{cccc}\toprule
                                    & cn125          & cn132                  & laptop    \\\midrule
        CPU                         & Xeon E-2378    & Epyc 7313 $(\times 2)$ & i7-12700H \\
        Cores                       & 8              & 16 $(\times 2)$        & 8         \\
        Vector extension            & AVX-512        & AVX2                   & AVX2      \\
        Base--Boost frequency (GHz) & 2.6--4.8       & 3.0--3.7               & 2.3--4.7  \\
        Theoretic bandwidth (GB/s)  & 51.2           & 204.8                  & 76.8      \\
        STREAM (GB/s)               & 41             & 140                    & 52        \\
    \bottomrule\end{tabular}
    \caption{Theoretical and practical capabilities of test systems}
    \label{tab:system}
\end{table*}

\subsection{The Datasets}\label{subsec:datasets}

The performance of the Quickhull algorithm depends on the input distribution, so
we briefly describe what makes the three datasets of PBBS interesting. This
includes the structure of the recursion, and the number of bytes that must be
moved. We compute the number of bytes as follows. Assuming a double is $8$
bytes, this is $8n$ bytes for finding the left-most and right-most points. At
each recursive call, we have $8(n + |S_1| + |S_2|)$ bytes for the partition, and
$8|CH(S_2)|$ bytes to place $CH(S_1)$ and $CH(S_2)$ next to each other in
memory. This is \qty{5.6}{\giga\byte}, \qty{73}{\giga\byte},
\qty{7.2}{\giga\byte} for Kuzmin, Circle, and Disk respectively.

\begin{figure}[ht]
    \centering
    \includegraphics[width=0.5\linewidth]{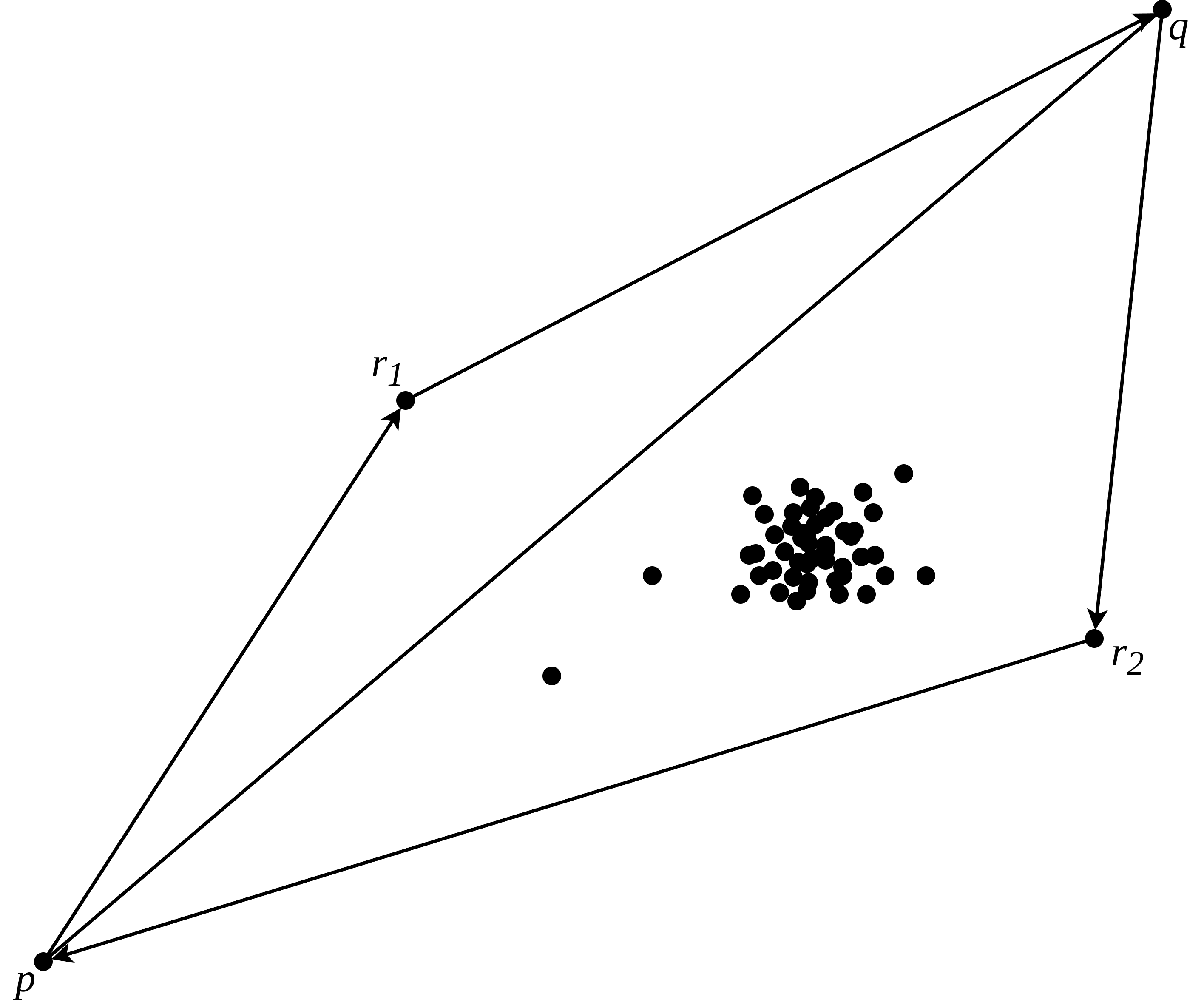}
    \caption{The Kuzmin dataset of $10^8$ points. The first partition only moves
    $r_1$. The second partition eliminates all remaining points.}
    \label{fig:kuzmin}
\end{figure}

The first data set is drawn according to a Kuzmin distribution, which we
illustrate in Figure~\ref{fig:kuzmin}. We only have one recursive call, and the
recursion is very unbalanced.

The second data set consists of points sampled uniformly from a circle, so
points of the form $(\cos(\theta), \sin(\theta))$. A true circle is convex, so
all its points lie on the convex hull. However, the algorithm finds only
$10985400$ points ($\sim$\qty{11}{\percent}). This is not a mistake or a 
problem because a sampled circle locally resembles a line
up to the square root of the unit roundoff, and points on a line can be 
rejected from the convex hull. To be more precise, at 
$(\cos(\theta), \sin(\theta))$, the tangent 
line has equation
\[
\cos(\theta) x + \sin(\theta) y = 1
\]
The distance between $(\cos(\theta + \delta), \sin(\theta + \delta))$ and this
line is
\[
|\cos(\theta)\cos(\theta + \delta) + \sin(\theta + \delta)\sin(\theta) - 1| =
\cos(\delta) - 1 = O(\delta^2)
\]
The unit round-off $u$ is $2^{-53}$ for double precision, so points on a circle
that are approximately $\sqrt{u} \approx 10^{-8}$ apart, are indistinguishable
from a line and even slight round-off errors create concave points. Regardless,
the recursion is deep and balanced.

The last data set consists of points sampled uniformly on a disk, the interior
of a circle. The recursion is also balanced, but because we discard more points
the deeper recursion levels are less expensive compared to Circle.

\subsection{Metrics}

Presenting performance results of an algorithm and its implementation in
a meaningful way requires more than the runtime. The runtime alone tells us 
little about the quality of the implementation, as we cannot relate it to the 
capabilities of the machine. For this reason, we present
our results using two statistics.

First---to evaluate how close the wallclock time gets to the hardware's
peak---we compute the bandwidth, or the amount of data moved divided by time. We
split the $x$- and $y$-coordinates into two separate arrays. Assuming 64-bit
floating-point numbers are used, this is $8|P|$ bytes for finding the left-most
and right-most points. At each recursive call, we have $8(|P| + |S_1| + |S_2|)$
bytes for the partition, and $8|CH(S_2)|$ bytes to place $CH(S_1)$ and $CH(S_2)$
next to each other in memory. This results in \qtylist{5.6;73;7.2}{\giga\byte}
for Kuzmin, Circle, and Disk respectively. Second---we measure the energy 
consumption in Joules. 

\subsubsection{Methodology}

We use GCC 14.1.1. All experiments are run \qty{10} times, and we plot the
standard deviation with error bars. 

In an attempt to generalize the energy consumption results across different
machines, we subtract the `idle' power draw of the system under test from the
observed energy consumption. We define idle power draw as the baseline power
that is required to keep the hardware operational, as well as the power required
for the operating system and additional background tasks. After a five-minute
measurement period, where both systems were fully reserved to ensure that no
other tasks are running, the idle power draw of cn125 was found to be
\qty{3.7}{\watt}, and the idle power draw of cn132 was found to be
\qty{53.0}{\watt} for the first of the two CPUs, and \qty{50.8}{\watt} for the
second. The difference in power draw between these two nominally identical CPUs
is expected, as silicon quality varies greatly even between CPUs of the same
model~\cite{cpu-variation}. \textit{laptop} has an idle power draw of
\qty{5.4}{\watt}.

Intel and AMD both provide hardware counters that track the accumulated energy
consumption of their processors. Although initially based on estimation models,
these interfaces nowadays provide precise energy consumption
statistics~\cite{rapl}. Intel CPUs provide these counters since their Sandy
Bridge architecture (2011), whereas AMD started providing them with their Zen
architecture (2017). These counters track the energy consumption of the entire
CPU, including the uncore and DRAM controller. Although ideally we would also
measure the energy consumption of the memory bus and the DRAM itself, these
capabilities are not available to us.

\subsection{Differences with PBBS}

The PBBS implementation partitions an array of indices, instead of the points
themselves. Their parallelization strategy is to spawn threads on the recursive
call until a base case of $400$ points is reached. The points $r_1$, $r_2$ are
found in a separate pass.

The partition algorithm applies Hoare's algorithm~\cite{Hoare62} sequentially,
and is scan-based~\cite{Blelloch89} in parallel. Because of this, runtime
speedup and energy-efficiency improvements of PBBS are much greater when
comparing the parallel case against the sequential case than for \vqhull{},
which applies the same approach in both the sequential and parallel case.

Finally, \vqhull{} uses separate arrays for $x$- and $y$-coordi\-nates, whereas
PBBS uses a single array of data type:
\begin{lstlisting}
    struct {
        double x;
        double y;
    };
\end{lstlisting}

\subsection{Computational Performance Analysis}

\paragraph{Branching}

\begin{figure*}[ht]
    \centering
    \includegraphics[width=\linewidth]{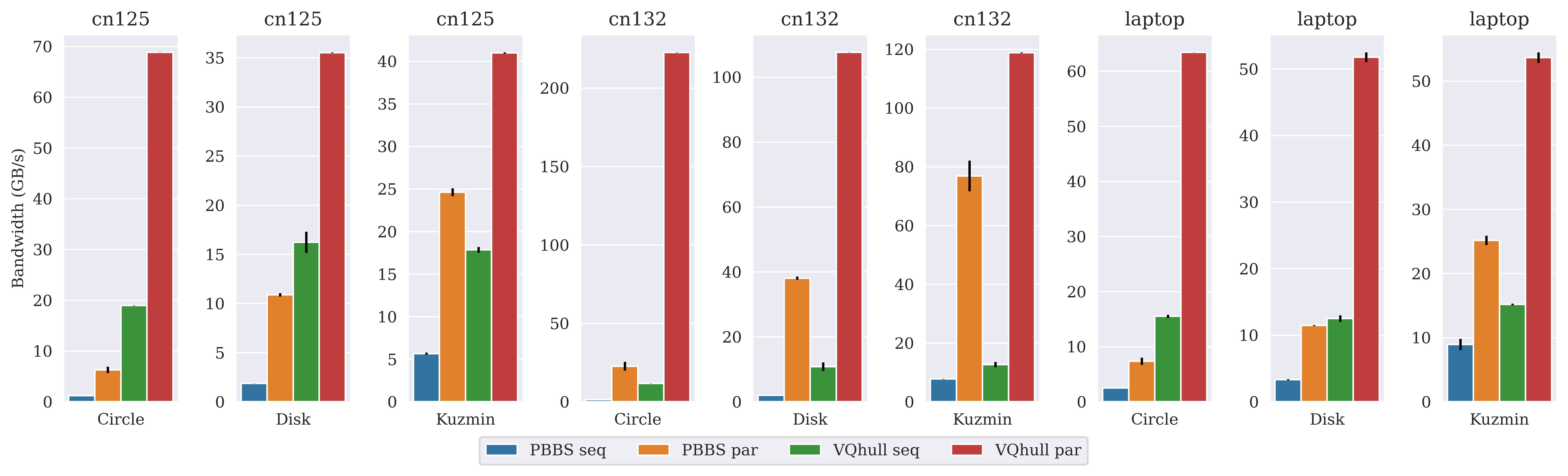}
    \caption{VQhull and PBBS bandwidth in GB/s.}
    \label{fig:bw}
\end{figure*}

For the Kuzmin dataset almost all points belong in either $S_1$ or in $S_2$, so
the branches of Hoare's partition algorithm are easy to predict. \vqhull{} on
the other hand only has one difficult branch every \qty{4} or \qty{8} elements.
This is reflected in Figure~\ref{fig:bw}, where the sequential PBBS
implementations differs as much as a factor \qty{6} between different inputs, 
while \vqhull{} differs by at most \qty{20}{\percent}.

\paragraph{Vectorization}

Even if branch misprediction and memory bandwidth are not an issue, so for
sequential Kuzmin, the \vqhull{} implementation is still
\qtyrange{1.5}{3.2}{\times} faster than the PBBS implementation, as SIMD
instructions can compute the tests \qty{4} or \qty{8} points at a time. We do
not get a speedup this large as the compression instruction is not
\qtyrange{4}{8}{\times} faster than the non-vectorized counterpart, and because
some SIMD instructions can be used within a single test as well.

\begin{figure*}[ht]
    \centering
    \includegraphics[width=\linewidth]{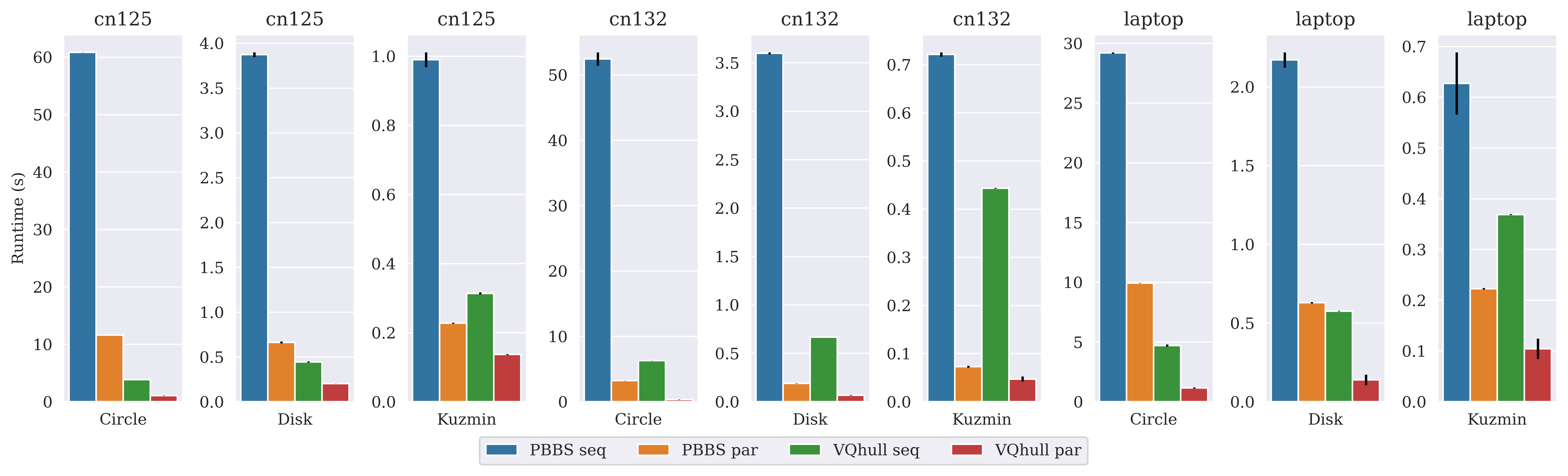}
    \caption{VQhull and PBBS runtime in seconds.}
    \label{fig:runtime}
\end{figure*}

\subsection{Memory Subsystem Analysis}

The weaker parallel performance of PBBS can be explained by the actual amount of 
traffic over the bus being higher than the minimum amount we use for our 
calculations. Finding $r_1$ and $r_2$ each take an additional extra pass over 
the data, which explains why \vqhull{} is faster. As the points are not 
permuted, only part of the cacheline is used effectively at deeper recursion 
levels of Disk and Circle. This explains why they get lower bandwidth than 
Kuzmin for PBBS.

For \vqhull{} we obtain \qtyrange{98}{100}{\percent} of STREAM's bandwidth for
Kuzmin on non-NUMA systems, and \qtyrange{85}{90}{\percent} for Disk. Disk's
performance may be lower because it is unpredictable from which side of the
array we read, making prefetching less effective. For Circle, we exceed the
RAM's bandwidth because significant work is done on subsets fitting in cache.

On cn132, a dual socket system with a NUMA architecture, we only obtain
\qtyrange{66}{80}{\percent} of STREAM. Profiling shows the first subset
extraction is done at \qty{100}{\percent} of STREAM, but deeper recursions are
significantly slower. This can be explained by suboptimal locality of threads
spawned in recursive calls, which we do not control.

\subsection{Energy Consumption Analysis}

Usually, going from the sequential implementation to the corresponding parallel
implementation decreases the total energy consumption. At first sight this seems
counter-intuitive, as in the end the same amount of work is being done by both
implementations, albeit with the work being distributed over multiple threads in
the parallel case. If anything, one might expect the additional energy overhead
related to scheduling to result in an increase in energy consumption. However,
CPUs typically run at a higher frequency for single-threaded workloads than for
parallel workloads. Because power draw is roughly proportional to cubic
frequency~\cite{cpu-power-frequency}, this explains why parallelizing a
workload, and consequently computing at a lower frequency, can decrease total
energy consumption.

The only two exceptions to this observation occur on cn125, for the Disk and
Kuzmin datasets. For these two cases, even though the runtime decreases going
from sequential \vqhull{} to the parallel implementation, the energy consumption
slightly increases. This highlights a case where an increase in energy
consumption, due to increased complexity of the implementation, as well as
additional operating system overhead, is not overcome by hardware efficiency
improvements resulting from a parallelized approach.

Furthermore, we see that even within a single implementation, a decrease in
runtime does not correlate to a decrease in energy consumption. Although
sequential PBBS is faster on cn132 than on cn125 for every dataset, it
conversely consumes more energy. Compared to \textit{laptop}, this is only the
case for the Kuzmin dataset. Whereas for Circle and Disk, \textit{laptop} is
both faster and more energy-efficient. This highlights that observing runtime
performance alone does not give the whole picture. And that selecting the
optimal implementation of an algorithm is highly context-dependent, depending
not only on the implementation itself but also on the hardware.

Finally, our experiments show that optimizing an implementation has gains beyond
improving runtime performance. Although parallel PBBS on average has a shorter
runtime than \vqhull{}, in all but one case sequential \vqhull{} results in
significantly lower energy consumption. This observation shows that decreasing
the overall amount of branch-misses and computational work, as well as making
more effective use of the available hardware and instruction set, allows for
substantial gains in energy-efficiency, even when runtime speedups are minimal.

\begin{figure*}[ht]
    \centering
    \includegraphics[width=\linewidth]{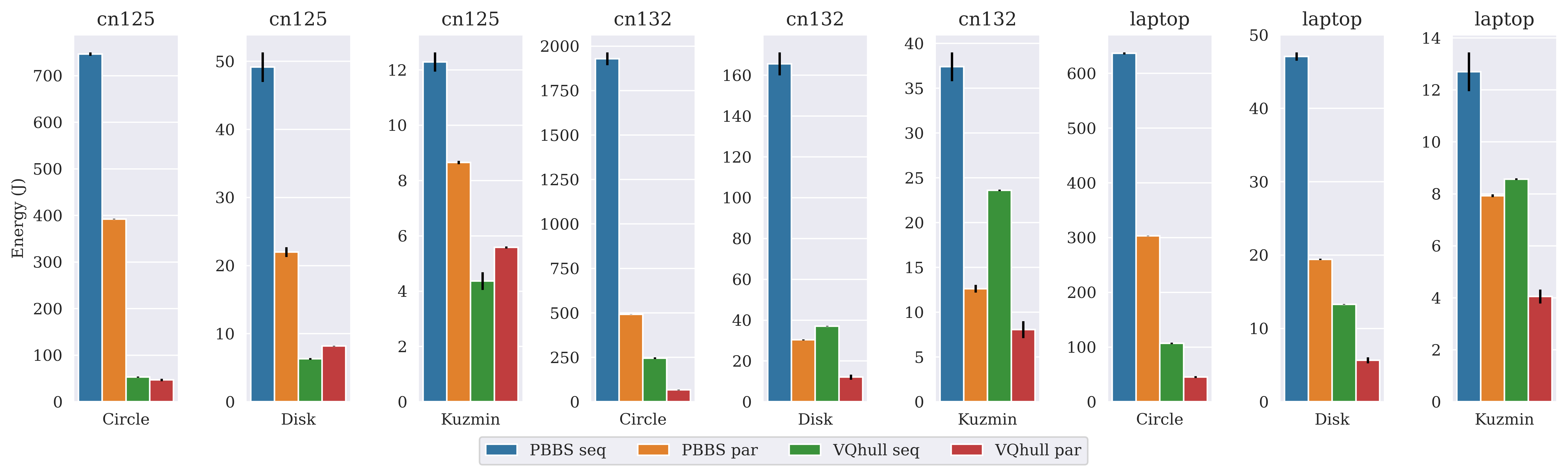}
    \caption{VQhull and PBBS energy consumption in Joules.}
    \label{fig:energy}
\end{figure*}

\section{Related and Future Work}\label{sec:related}

Finding planar convex hulls is a well-studied problem, there exist at least ten
algorithms for solving it \cite{Graham72, Jarvis73, Eddy77, Preparata77,
Bykat78, Akl78, Andrew79, Clarkson93, Barber96, Chan96}. Putting our work in
context also suggests several opportunities for future work.

The sequential subset extraction is a straightforward adaptation of the
vectorized partition algorithm used in QuickSort~\cite{Bramas17}.

\begin{figure}[ht]
    \centering
    \resizebox{0.5\linewidth}{!}{%
    \begin{tikzpicture}
        \draw (0, 0) circle (2);
        \draw (-2, 0) node[left] {$p_1$} --
              (-1.414, 1.414) node[above left] {$p_2$} --
              (0, 2) node[above] {$p_3$} --
              (1.414, 1.414) node[above right] {$p_4$} --
              (2, 0) node[right] {$p_5$} --
              (1.414, -1.414) node[below right] {$p_6$} --
              (0, -2) node[below] {$p_7$} --
              (-1.414, -1.414) node[below left] {$p_8$} --
              (-2, 0);
    \end{tikzpicture}}
    \caption{Points $p_1, \cdots, p_8$ can be found by minimum / maximum
    $x$-coordinates, $y$-coordinates, differences / sums of $x$ and $y$
    coordinates. Points within this octagon are not in the convex hull.}
    \label{fig:akl}
\end{figure}
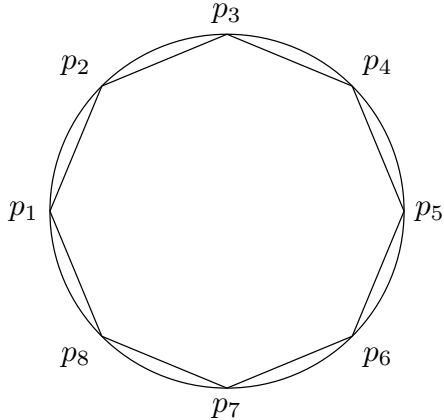

Our analysis shows the implementation \vqhull{} is limited by bandwidth, and
that it gets close to the theoretic maximum any implementation of Quickhull can
obtain. For this reason, significant speedups can only be obtained by
implementing an algorithm with less data movement than Quickhull. With this in
mind, the Akl-Toussaint heuristic~\cite{Akl78} would be an interesting
opportunity for future work. In the case of Disk, this heuristic lets us
identify $p_1, \cdots, p_8$ of Figure~\ref{fig:akl} in one pass. This octagon
has area $2\sqrt{2}$, which is approximately \qty{90}{\percent} of the disk, and
these points can be eliminated. Points $u$ satisfy $\orient(p_i, u, p_{i + 1
\,\text{mod}\, 8}) > 0$ for precisely one $i$ if they are between the circle and
line segment $\overrightarrow{p_i p_{i + 1 \,\text{mod}\, 8}}$, or none if $u$
is inside the octagon. If we can partition the input into these eight areas in
one pass as well, we need only $1.6 + 1.6 + 0.16 = 3.36$ GB to eliminate
\qty{90}{\percent} of points. \vqhull{} gets to the same points in \qty{3}
levels of recursion, which takes \qty{6.5}{\giga\byte}. So a better algorithm
can cut the required bandwidth almost in half.

There is some work on computing convex hulls on Graphic Processor Units (GPUs)
\cite{Srungarapu11}. They report a $16\times$ speedup compared to Qhull, which
is not competitive with \vqhull{}. However, single-threaded performance has not
increased nearly at the rate of GPU performance, so this may have changed. Their
implementation is not publicly available.

\section{Conclusion}

We present \vqhull{}, a parallel algorithm for \qhull{} Our implementation shows
an improvement of up to \qty{16}{\times} compared to the state-of-the-art
sequential implementation. By focussing on the memory subsystem instead of
computational parallelism, also the multithreaded implementation shows an
improvement of up to \qty{11}{\times} improvement over the state-of-the-art
parallel implementation.

Our implementations achieve \qtyrange{66}{100}{\percent} of the architecture's
peak bandwidth, showing the potential for further improvements is
\qtyrange{0}{50}{\percent}. \vqhull{} does best on systems with a single CPU,
and an instruction set that supports a compression instruction. A system with
multiple CPUs and a non-uniform memory architecture shows significant
performance degradation for nested OpenMP parallelism, providing an opportunity
for future work.

This algorithm also serves as a case study on engineering algorithms for
energy-efficiency. We show that parallelizing programs is beneficial to
energy-efficiency, resulting from a decrease in clock frequency. Energy savings
on specific architectures are mostly in line with runtime improvements, but even
when runtime savings are minimal, a well-thought-out implementation can still
result in significant energy savings.

\bibliographystyle{plain}
\bibliography{paper.bib}

\end{document}